\documentclass[
 reprint,
 amsmath,amssymb,
 aps,
 prl,
 longbibliography
]{revtex4-2}

\usepackage{graphicx}
\usepackage{dcolumn}
\usepackage{bm}
\usepackage{physics}
\usepackage{float}
\usepackage{color}
\usepackage{hyperref}
\hypersetup{
     colorlinks   = true,
     linkcolor    = blue,
     citecolor    = blue,
     urlcolor     = blue
}

\begin{document}

\title{Thermodynamic Electric Toroidal Dipole and Intrinsic Longitudinal Spin Transport
}

\author{Taisei Yamanaka}
 \email{yamanaka.taisei.i3@elms.hokudai.ac.jp}
\author{Takumi Sato}
 \email{sato@phys.sci.hokudai.ac.jp}
\author{Satoru Hayami}
 \email{hayami@sci.hokudai.ac.jp}
\affiliation{
Graduate School of Science, Hokkaido University, Sapporo 060-0810, Japan
}

\date{\today}

\begin{abstract} 
Electric toroidal dipoles (ETDs) characterize ferroaxial order, yet their bulk definition in periodic crystals has remained elusive because conventional multipole operators involve the ill-defined position operator. 
Here we formulate a thermodynamic ETD by coupling a spatially varying electric field to the relativistic spin-induced electric polarization. 
The resulting expression is gauge invariant and provides a bulk order parameter for ferroaxial phases. 
We further establish a direct relation between the chemical-potential derivative of the ETD and the intrinsic longitudinal spin conductivity in insulating systems. 
To demonstrate the formulation, we construct a minimal ferroaxial extension of the Kane--Mele model. 
The ETD becomes finite exclusively in the ferroaxial phase and is strongly enhanced near a small band gap, accompanied by a sizable longitudinal spin current. 
Our results establish a thermodynamic theory of ETDs in crystalline solids and identify the longitudinal spin conductivity as a direct transport manifestation of ferroaxial order.
\end{abstract}

\maketitle
Ferroaxial order has recently emerged as an unconventional symmetry-breaking phase that preserves both spatial inversion and time-reversal symmetries while breaking the in-plane mirror symmetry~\cite{Hlinka_PhysRevLett.113.165502, Hlinka_PhysRevLett.116.177602, zeng2025photo, Day_PhysRevLett.134.016401}.
Although it does not couple directly to electromagnetic fields, it gives rise to a variety of unconventional phenomena~\cite{cheong2021permutable, kirikoshi2023rotational}, including longitudinal spin-current generation~\cite{hayami2022electric, roy2022unconventional}, antisymmetric thermopolarization~\cite{nasu2022antisymmetric}, nonlinear transverse magnetization~\cite{inda2023nonlinear,du2026electric}, unconventional Hall effects~\cite{hayami2023unconventional}, and nonlinear Edelstein effects~\cite{m4k3-mttk}.
The recent discovery of ferroaxial order in a growing number of materials, including Cu$_3$Nb$_2$O$_8$~\cite{johnson2011prl_Cu3Nb2O8}, CaMn$_7$O$_{12}$~\cite{johnson2012prl_CaMn7O12}, RbFe(MoO$_4$)$_2$~\cite{jin2020observation, Hayashida_PhysRevMaterials.5.124409}, NiTiO$_3$~\cite{hayashida2020visualization, Hayashida_PhysRevMaterials.5.124409, yokota2022three, Bhowal_PhysRevResearch.6.043141}, BaCoSiO$_4$~\cite{xu2022prb_BaCoSiO4}, Ca$_5$Ir$_3$O$_{12}$~\cite{Hasegawa_doi:10.7566/JPSJ.89.054602, hanate2021first, hayami2023cluster, hanate2023space}, K$_2$Zr(PO$_4$)$_2$~\cite{yamagishi2023ferroaxial, Bhowal_PhysRevResearch.6.043141, j6tr-ggn9}, Na$_2$Hf(BO$_3$)$_2$~\cite{nagai2023chemicalSwitching}, Na-superionic conductors~\cite{nagai2023chemical}, MnTiO$_3$~\cite{Sekine_PhysRevMaterials.8.064406}, 1$T$-TaS$_2$~\cite{Luo_PhysRevLett.127.126401, liu2023electrical}, and Zn$_3$(V$_2$O$_7$)(OH)$_2\cdot$2H$_2$O~\cite{10.1021/jacs.5c19407}, has further stimulated interest in its microscopic characterization.
 
Within the multipole description of electronic order, ferroaxial order is characterized by the electric toroidal dipole (ETD)~\cite{hayami2024unified}.
The ETD is classically represented by a vortex of electric polarization~\cite{Dubovik1975MultipoleEI,DUBOVIK1990145},
\begin{align}
  \vb*{G}^{\text(c)} \propto \sum_i \vb*{r}_i \times \vb*{P}_i ,
\end{align}
where, $\vb*{r}_i$ and $\vb*{P}_i$ denote the position and electric polarization at site $i$, respectively.
In periodic crystals, however, the conventional expression cannot be directly evaluated because it explicitly involves the position operator $\bm{r}$, which is ill defined in the Bloch representation.
An analogous difficulty arises for the bulk electric polarization and is resolved by the modern theory of polarization~\cite{king1993theory,vanderbilt1993electric,resta1994macroscopic}, in which polarization is expressed through the Berry phase of Bloch wave functions and is uniquely determined only modulo the lattice points. 
Generalizing this construction to higher multipole moments is nontrivial because their conventional definitions involve higher powers of the position operator and do not, in general, admit a simple Berry-phase representation.

Significant progress toward overcoming this difficulty has recently been achieved through the thermodynamic theory of multipole moments, in which bulk multipoles are formulated as coefficients in the free-energy expansion with respect to spatially varying external fields~\cite{shi2007quantum,gao2018microscopic,shitade2018theory,gao2018orbital,shitade2019theory,daido2020thermodynamic,oike2025thermodynamic,sato2026, sato2026orbitalmagneticoctupolecrystalline, yamanaka2026magnetic}.
This framework provides gauge-invariant expressions for a wide variety of multipole moments in periodic crystals.
The ETD, however, has remained an outstanding exception.
From symmetry considerations, the ETD should be identified with the antisymmetric component of the thermodynamic electric quadrupole.
In contrast, the thermodynamic electric quadrupole derived within the existing formalism is intrinsically symmetric~\cite{daido2020thermodynamic}, thereby precluding a thermodynamic description of the ETD.
Resolving this inconsistency is essential for establishing a bulk theory of ferroaxial order.

In this work, we resolve this longstanding problem by incorporating the relativistic spin-induced electric polarization~\cite{katsura2005spin, hoshino2023spin, hayami2024analysis} into the thermodynamic multipole formalism.
We formulate a gauge-invariant thermodynamic ETD as the free-energy response to a spatially varying electric field coupled to the relativistic electric polarization.
Furthermore, we establish that the chemical-potential derivative of the ETD is directly related to the longitudinal spin conductivity in insulating systems, providing a direct connection between the thermodynamic order parameter and intrinsic spin transport.
To demonstrate the theory, we construct a minimal ferroaxial extension of the Kane--Mele model and show that the thermodynamic ETD correctly characterizes ferroaxial order.
The ETD becomes finite only in the ferroaxial phase and is strongly enhanced near a small band gap, accompanied by a sizable longitudinal spin conductivity.

To formulate the thermodynamic ETD, we incorporate the relativistic spin-induced electric polarization into the free-energy response to a spatially varying electric field.
Throughout this paper, we use the units of $k_{\text{B}} = c = \hbar = 1$, where $k_{\text{B}}$ is the Boltzmann constant and $c$ is the speed of light.
Following the derivation of the thermodynamic multipole~\cite{shi2007quantum,daido2020thermodynamic},
we consider
\begin{align}\label{eq:dF}
  dF
  =
  -S\,dT
  -N\,d\mu
  -P_i\,dE_i
  -Q_{ij}\,d(\partial_jE_i)
  -\cdots,
\end{align}
where $S$, $T$, $N$, and $\mu$ denote the entropy, temperature, particle number, and chemical potential, respectively. 
The coefficients $P_i$ and $Q_{ij}$ are identified as the electric dipole and quadrupole moments, respectively, associated with the response to a spatially varying electric field.

Relativistic effects generate a spin-dependent contribution to the electric polarization~\cite{Wang_PhysRevLett.96.066601, hoshino2023spin},
\begin{align}
  \vb*{P}
  &=
  \vb*{P}^{\mathrm{charge}}
  +
  \vb*{P}^{\mathrm{spin}},
  \nonumber\\
  \vb*{P}^{\mathrm{charge}}
  &=
  -e\vb*{r},
  \qquad
  \vb*{P}^{\mathrm{spin}}
  =
  \xi\,\vb*{s}\times\vb*{v},
\end{align}
where $e$ is the elementary charge and
$\xi=\hbar/(2mc^2)$ characterizes the relativistic correction.
In the following, we focus on the spin-induced polarization;
the charge contribution has been discussed within the conventional thermodynamic electric-quadrupole formalism~\cite{daido2020thermodynamic}.
Accordingly, we define
\begin{align}
  Q_{ij} := -\left( \frac{\partial F}{\partial [\partial_{r_j}E_i]} \right).
\end{align}
Unlike the conventional formulation~\cite{daido2020thermodynamic}, this definition allows antisymmetric components, thereby enabling a thermodynamic definition of the ETD.

The Maxwell relation associated with Eq.~\eqref{eq:dF} gives
\begin{align}
  \frac{\partial Q_{ij}}{\partial \mu} = \frac{\partial N}{\partial [\partial_{r_j}E_i]},
\end{align}
where the right-hand side can be evaluated from the long-wavelength expansion of the density--polarization correlation function~\cite{oike2025thermodynamic,sato2026orbitalmagneticoctupolecrystalline},
\begin{align}
  \chi_{N, P^{\mathrm{spin}}_i}(\vb*{q}, \omega) &= \xi \int \frac{d^dk}{(2\pi)^d} \sum_{n,m}^{} \bra{n\vb*{k}_-}\ket{m\vb*{k}_+} \nonumber\\
  &\times \bra{m\vb*{k}_+}  \frac{1}{2}\left(\hat{\tilde{r}}_k (\vb*{k}_+) + \hat{\tilde{r}}_k (\vb*{k}_-) \right) \ket{n\vb*{k}_-} \nonumber\\
  &\times \frac{f_{n\vb*{k}_-} - f_{m\vb*{k}_+}}{\epsilon_{n\vb*{k}_-} - \epsilon_{m\vb*{k}_+} + \omega + i\delta},
\end{align}
which describes the linear response $\delta N (\vb*{q}, \omega) =  \chi_{N, P^{\mathrm{spin}}_i}(\vb*{q}, \omega)E_i(\vb*{q}, \omega)$ with the wave vector $\bm{q}$ and the frequency $\omega$.
Here, $\hat{\mathcal{H}}_{\vb*{k}}$ is the Bloch Hamiltonian with eigenvalues $\epsilon_{n\bm{k}}$ for eigenstates $\ket{n\bm{k}}$ ($n$ labels the band index and $\vb*{k}$ is the crystal momentum), $f_{n\vb*{k}}= (1 + e^{\beta(\epsilon_{n\vb*{k}} - \mu)})^{-1}$ is the Fermi distribution function, $\hat{\tilde{r}}_i (\vb*{k}) = (1/2)\epsilon_{iab}\left\{ \hat{s}_a, \hat{v}_b (\vb*{k})\right\}_+$ with $\hat{v}_j (\vb*{k}) = \partial_{k_j} \hat{\mathcal{H}}_{\vb*{k}}$ ($\partial_{k_j}= \partial / \partial k_{j}$), and $\vb*{k}_\pm = \vb*{k}\pm \vb*{q}/2$.

Using the response function above, the electric quadrupole is obtained as
\begin{align}
  \frac{\partial Q_{ij}}{\partial \mu} &= -i\lim_{\vb*{q} \to 0} \partial_{q_j} \lim_{\delta \to 0} \chi_{N, \tilde{P}_i}(\vb*{q}, 0) .
\end{align}
Integrating over the chemical potential yields
\begin{align}\label{eq:EQ}
  Q_{ij} 
  &= \xi\int_{}^{}\frac{d^dk}{(2\pi)^d} \sum_{n}^{} 
  \Bigg[ \Omega^{\tilde{r}_i, v_j}_{n\vb*{k}} \mathcal{G}_{n\vb*{k}} + m^{\tilde{r}_i, v_j}_{n\vb*{k}}f_{n\vb*{k}} \Bigg],
\end{align}
where $\mathcal{G}_{n\vb*{k}} = -T\log\left\{1 + e^{-\left(\epsilon_{n\vb*{k}} - \mu\right)/T}\right\}$ is the grandpotential density, and 
\begin{align}\label{eq:omega}
  \Omega^{\tilde{r}_i, v_j}_{n} &= -\sum_{m}^{\neq n}\frac{2\Im\left[ 
    \bra{n\vb*{k}} \hat{\tilde{r}}_i (\vb*{k})\ket{m\vb*{k}}
    \bra{m\vb*{k}} \hat{v}_j(\vb*{k}) \ket{n\vb*{k}} \right]}{(\epsilon_{n\vb*{k}} - \epsilon_{m\vb*{k}})^2}, \nonumber\\
  m^{\tilde{r}_i, v_j}_{n} &= \sum_{m}^{\neq n}\frac{\Im\left[
    \bra{n\vb*{k}} \hat{\tilde{r}}_i (\vb*{k}) \ket{m\vb*{k}}
    \bra{m\vb*{k}} \hat{v}_j(\vb*{k}) \ket{n\vb*{k}} \right]}{\epsilon_{n\vb*{k}} - \epsilon_{m\vb*{k}}}.
\end{align}
Here, $\Omega^{\tilde r_i,v_j}_{n\vb{k}}$ and $m^{\tilde r_i,v_j}_{n\vb{k}}$ represent the generalized Berry-curvature and orbital-moment contributions, respectively.
A detailed derivation is provided in the Supplemental Material (SM)~\cite{suppl}.

Crucially, the tensor $Q_{ij}$ obtained from the spin-induced polarization generally contains an antisymmetric component.
This enables us to define the thermodynamic ETD as $G_k = (1/2) \epsilon_{ijk} Q_{ij}.$
In addition, the electric-monopole (EM) component is defined by the trace part of $Q_{ij}$ as $Q_0 = (1/3) Q_{ii}.$
$G_k$ therefore provides a bulk thermodynamic definition of the ETD applicable to periodic crystals.
In the static situation considered here, $\nabla \times \vb*{E} = 0$; therefore, the ETD, which is conjugate to this quantity, does not contribute to the free energy.

The thermodynamic ETD is directly connected to an experimentally accessible transport response.
Specifically, we establish a direct relation between its chemical-potential derivative and the intrinsic spin conductivity.
Within the Kubo formalism, the intrinsic spin conductivity is given by~\cite{sinova2004universal,mook2020origin}
\begin{align}
  \sigma^{i}_{jk} = e&\int_{}^{}\frac{d^dk}{(2\pi)^d} \sum_{n, m}^{n \neq m} \frac{f_{n\vb*{k}} - f_{m\vb*{k}}}{(\epsilon_{n\vb*{k}} - \epsilon_{m\vb*{k}})^2} \nonumber\\
  &\times \Im\left[ 
    \bra{n\vb*{k}} \hat{j}^i_j(\vb*{k}) \ket{m\vb*{k}}
    \bra{m\vb*{k}} \hat{v}_k(\vb*{k}) \ket{n\vb*{k}} \right],
  \label{eq:sc}
\end{align}
where, $\hat{j}^i_j(\vb*{k}) := \frac{1}{2}\left\{ \hat{s}_i, \hat{v}_j(\vb*{k}) \right\}_+$ and $\sigma^{i}_{jk}$ is the response of the spin current to an electric field: $j^{i}_j = \sigma^{i}_{jk} E_k$.
For an insulator at zero temperature, Eq.~\eqref{eq:sc} reduces to
\begin{align}
   \sigma^{i}_{jk} = e&\int_{}^{}\frac{d^dk}{(2\pi)^d} \sum_{n}^{\text{occ}} \sum_{m}^{n \neq m}\frac{2\Im\left[ 
    \bra{n\vb*{k}} \hat{j}^i_j(\vb*{k}) \ket{m\vb*{k}}
   \bra{m\vb*{k}} \hat{v}_k(\vb*{k}) \ket{n\vb*{k}} \right]}{(\epsilon_{n\vb*{k}} - \epsilon_{m\vb*{k}})^2}.
\end{align}
Here, $\sum_{n}^{\text{occ}}$ denotes summation over all occupied bands.
Comparison with the chemical-potential derivative of Eq.~\eqref{eq:EQ} immediately yields
\begin{align}\label{eq:streda1}
  e\frac{\partial Q_{ij}}{\partial \mu} = -\xi \epsilon_{iab} \sigma^{a}_{bj}.
\end{align}
Taking the antisymmetric component of Eq.~\eqref{eq:streda1}, we obtain
\begin{align}\label{eq:streda2}
  e\frac{\partial G_k}{\partial \mu} = -\frac{1}{2}\xi \epsilon_{ijk}\epsilon_{iab} \sigma^{a}_{bj} = -\frac{1}{2}\xi (\sigma^j_{kj} - \sigma^k_{jj}).
\end{align}
Equation~\eqref{eq:streda2} is a central result of this work: the chemical-potential derivative of the thermodynamic ETD is directly related to a longitudinal component of the spin-conductivity tensor.

In particular, for a two-dimensional system in the $xy$ plane with a ferroaxial moment along $z$, Eq.~\eqref{eq:streda2} reduces to
\begin{align}
  e\frac{\partial G_z}{\partial \mu} = \frac{1}{2}\xi\left(\sigma^z_{xx} + \sigma^z_{yy} \right),
  \label{eq:Gz_sigma}
\end{align}
establishing a direct thermodynamic relation between the ETD and the longitudinal spin conductivity~\cite{hayami2022electric}.
For comparison, the EM component satisfies
\begin{align}\label{eq:Q0_sigma}
  e\frac{\partial Q_0}{\partial \mu} = -\frac{1}{3}\xi \epsilon_{iab} \sigma^{a}_{bi} = -\frac{1}{3} \xi \left( \sigma^z_{xy} - \sigma^z_{yx}\right),
\end{align}
which is associated with the spin Hall conductivity~\cite{murakami2003dissipationless, sinova2004universal}.

\begin{figure}[htbp]
\includegraphics[width=0.9\linewidth]{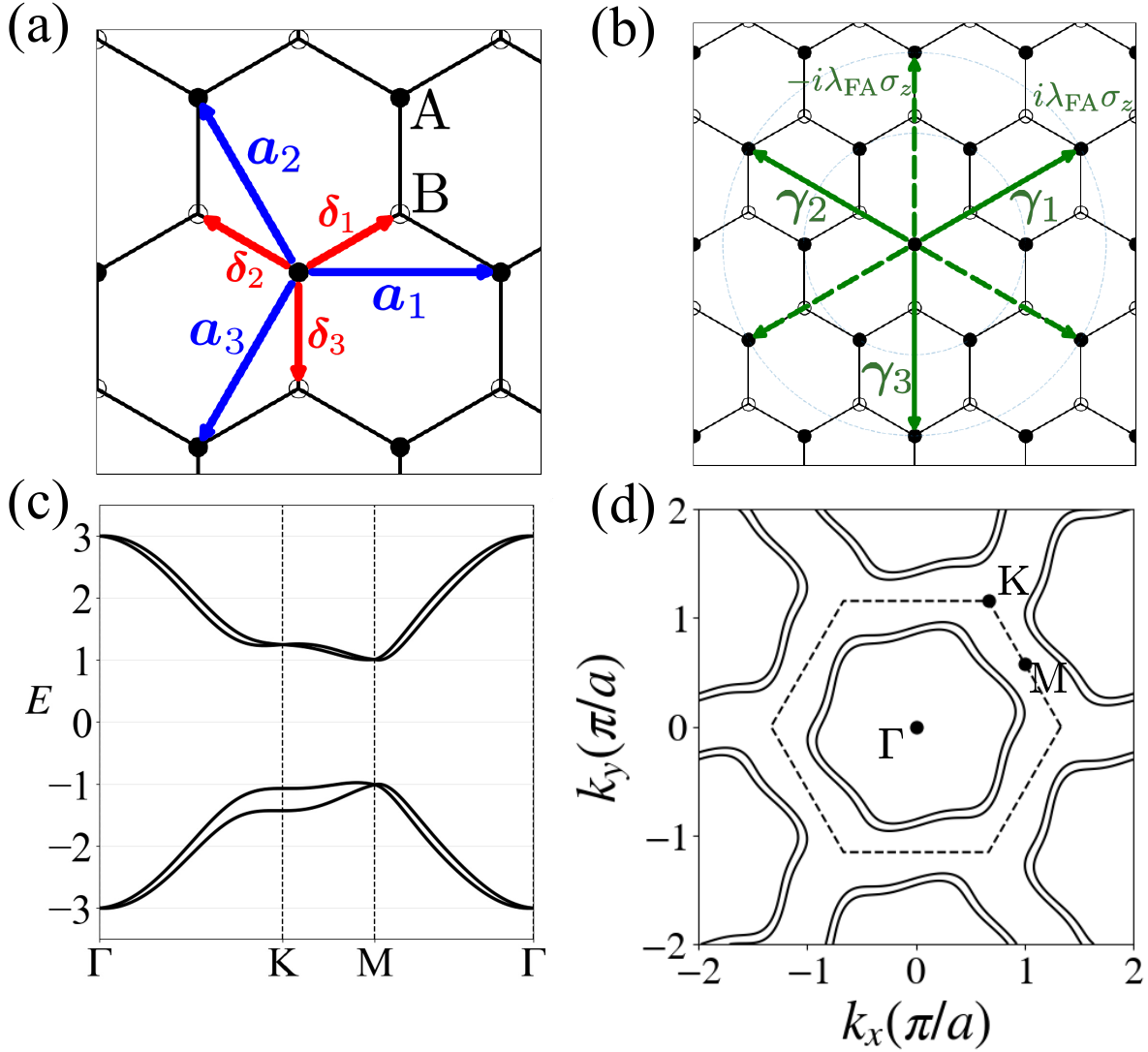}
\caption{
(a) Honeycomb structure
with the nearest-neighbor and next-nearest-neighbor vectors $\bm{\delta}_i$ and $\bm{a}_i$.
(b) Schematic representation of the ferroaxial hopping $\mathcal{H}_{\text{FA}}$.
(c) Band structure for $\lambda_{\text{FA}} = 0.1$.
The high-symmetry points are $\Gamma = (0, 0), \mathrm{K} = (0.5, 0), \mathrm{M} = (0.5, 0.5)$ in units of the reciprocal lattice vectors.
(d) Constant-energy contour at $E=1.5$ for $\lambda_{\mathrm{FA}}=0.1$.
The hexagon represents the first Brillouin zone.
}
\label{fig:model} 
\end{figure}

To demonstrate the formulation and elucidate the behavior of the thermodynamic ETD, we construct a minimal ferroaxial extension of the Kane--Mele model in the honeycomb structure.
We start from the Kane--Mele Hamiltonian~\cite{kane2005quantum,kane2005z}, including Rashba spin--orbit coupling,
\begin{align}
  \mathcal{H}_{\text{KM}} &= \mathcal{H}_{\text{t}} + \mathcal{H}_{\text{SO}} + \mathcal{H}_{\text{R}}, \nonumber \\
  \mathcal{H}_{\text{t}} &= -t\sum_{i}^{} \sum_{\sigma}^{}e^{-i\vb*{k} \cdot \vb*{\delta}_i} c^\dag_{\text{A} \sigma \vb*{k}}c_{\text{B} \sigma \vb*{k}} + \text{h.c.}, \nonumber\\
  \mathcal{H}_{\text{SO}} &= i\lambda_{\text{SO}} \sum_{i}^{} \sum_{\sigma\sigma'}^{} e^{-i\vb*{k} \cdot \vb*{a}_i} (\sigma_z)_{\sigma\sigma'} \nonumber\\
  \begin{split}
      &\times \left(c^\dag_{\text{A} \sigma \vb*{k}}c_{\text{A} \sigma' \vb*{k}} - c^\dag_{\text{B} \sigma \vb*{k}}c_{\text{B} \sigma' \vb*{k}}\right) + \text{h.c.},
  \end{split}
   \nonumber\\
  \mathcal{H}_{\text{R}} &= i\lambda_{\text{R}} \sum_i \sum_{\sigma\sigma'}^{} (\vb*{\sigma}_{\sigma\sigma'} \times \hat{\bm{\delta}}_i)_z e^{-i\bm{k}\cdot\bm{\delta}_i}c^\dag_{\text{A}\sigma\vb*{k}}c_{\text{B}\sigma'\vb*{k}} + \text{h.c.} ,
\end{align}
where $c^\dag_{\text{A}\sigma} (c_{\text{A}\sigma})$ is the creation (annihilation) operator of an electron with spin $\sigma$, at sublattice A or B.
The vectors $\vb*{\delta}_i$ and $\vb*{a}_i$ denote the nearest- and next-nearest-neighbor bonds, respectively [Fig.~\ref{fig:model}(a)], and $\hat{\vb*\delta}_i=\vb*\delta_i/|\vb*\delta_i|$.
In this model, both $G_z$ and the longitudinal spin conductivity vanish when $\lambda_{\text{R}} = 0$~\cite{suppl}.
The Kane--Mele model possesses the vertical mirror symmetries of the honeycomb structure. 
To construct a minimal model for ferroaxial order, we introduce the following symmetry-allowed hopping, which breaks the vertical mirror symmetries,
\begin{align}
  \mathcal{H}_{\text{FA}} &= i\lambda_{\text{FA}} \sum_{i}^{} \sum_{\sigma\sigma'}^{} e^{-i\vb*{k} \cdot \vb*{\gamma}_i} (\sigma_z)_{\sigma\sigma'} \nonumber\\
  \begin{split}
      &\times \left(c^\dag_{\text{A}\sigma\vb*{k}}c_{\text{A}\sigma'\vb*{k}} - c^\dag_{\text{B}\sigma\vb*{k}}c_{\text{B}\sigma'\vb*{k}}\right) + \text{h.c.}.
  \end{split}
\end{align}
This term describes an imaginary spin-dependent fifth-neighbor hopping, as illustrated in Fig.~\ref{fig:model}(b).
It breaks the vertical mirror symmetries, thereby allowing a finite ETD $G_z$.
We therefore identify $\lambda_{\mathrm{FA}}$ as the control parameter of ferroaxial order and consider $\mathcal{H} = \mathcal{H}_{\text{KM}} + \mathcal{H}_{\text{FA}}$ as the total Hamiltonian.
We set the lattice constant to $a=1$ and take $t=1$, $\lambda_{\text{SO}}=0.24$ and $\lambda_{\text{R}}=0.06$.

Figure~\ref{fig:model}(c) shows the band structure of the ferroaxial Kane--Mele model for $\lambda_{\mathrm{FA}}=0.1$.
The constant-energy contour in Fig.~\ref{fig:model}(d) exhibits a characteristic mirror-asymmetric deformation induced by the ferroaxial order, consistent with the broken vertical mirror symmetries.

\begin{figure}[htbp]
\includegraphics[width=\linewidth]{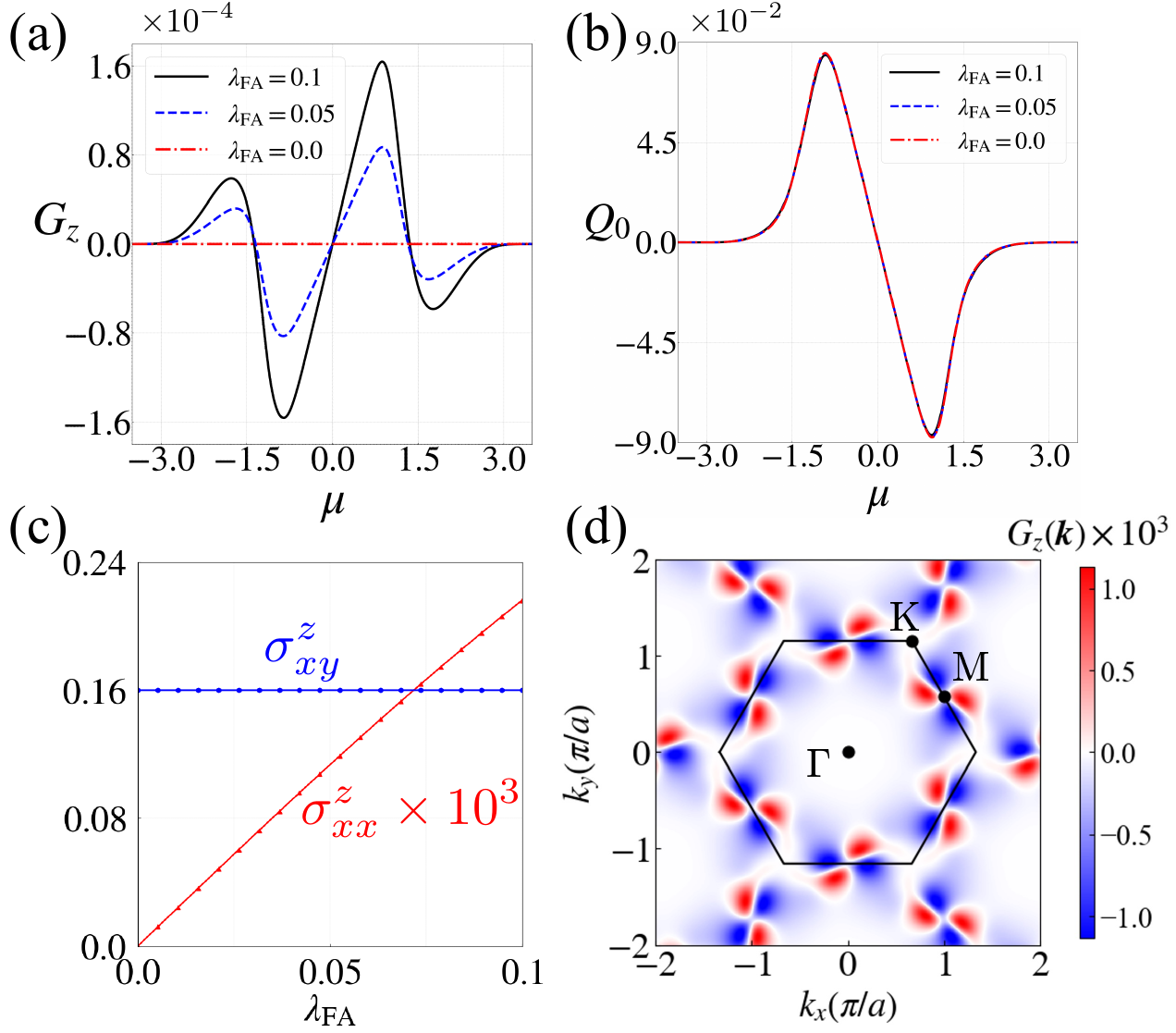}
\caption{
Chemical potential dependence of thermodynamic ETD $G_z$ and EM $Q_0$ for $\lambda_{\text{FA}} = 0.0$, $0.05$, $0.1$ at $T = 0.1$.
(c) $\lambda_{\text{FA}}$ dependence of $\sigma^z_{xy}$ and $\sigma^z_{xx}$ in the insulating regime at $T = 0.01$.
(d) Momentum-resolved contribution $G_z(\vb*{k})$ at $T = 0.01$ and $\mu = -0.5$.
The hexagon represents first Brillouin zone.
}
\label{fig:mu_dep} 
\end{figure}

We next evaluate the thermodynamic ETD using Eq.~\eqref{eq:EQ}. 
Figures~\ref{fig:mu_dep}(a) and (b) show the chemical-potential dependence of $G_z$ and $Q_0$ for several values of $\lambda_{\mathrm{FA}}$ at $T=0.1$, calculated using a $1024^2$ momentum mesh.
For $\lambda_{\mathrm{FA}}=0$, $G_z$ vanishes, whereas it becomes finite once ferroaxial order is introduced and increases in magnitude with $\lambda_{\mathrm{FA}}$.
In contrast, $Q_0$ remains essentially unchanged.
These results demonstrate that the thermodynamic $G_z$ selectively captures the ferroaxial symmetry breaking.
Figure~\ref{fig:mu_dep}(c) show the $\lambda_{\mathrm{FA}}$ dependence of $\sigma^z_{xx}$ and $\sigma^z_{xy}$ at $T=0.01$ in the insulating regime.
We find that the numerical results satisfy Eqs.~\eqref{eq:Gz_sigma} and ~\eqref{eq:Q0_sigma}.

A pronounced enhancement of $G_z$ appears around $\mu\simeq1$.
Its origin can be traced to the small interband energy separation near the M point, since the geometric contribution in Eq.~\eqref{eq:omega} is enhanced when the interband energy difference in the denominator becomes small.
To visualize its momentum-space structure, we define
$G_z(\vb*{k})$ as $G_z = \int_{}^{}\frac{d^dk}{(2\pi)^d} G_z(\vb*{k})$ and plot $G_z(\vb*{k})$ in Fig.~\ref{fig:mu_dep}(d).
The dominant contribution is concentrated around the M points, whereas the regions near $\Gamma$ and K contribute much less.
See the SM for a more detailed discussion of the contribution from the M point~\cite{suppl}.

\begin{figure}[htbp]
\includegraphics[width=\linewidth]{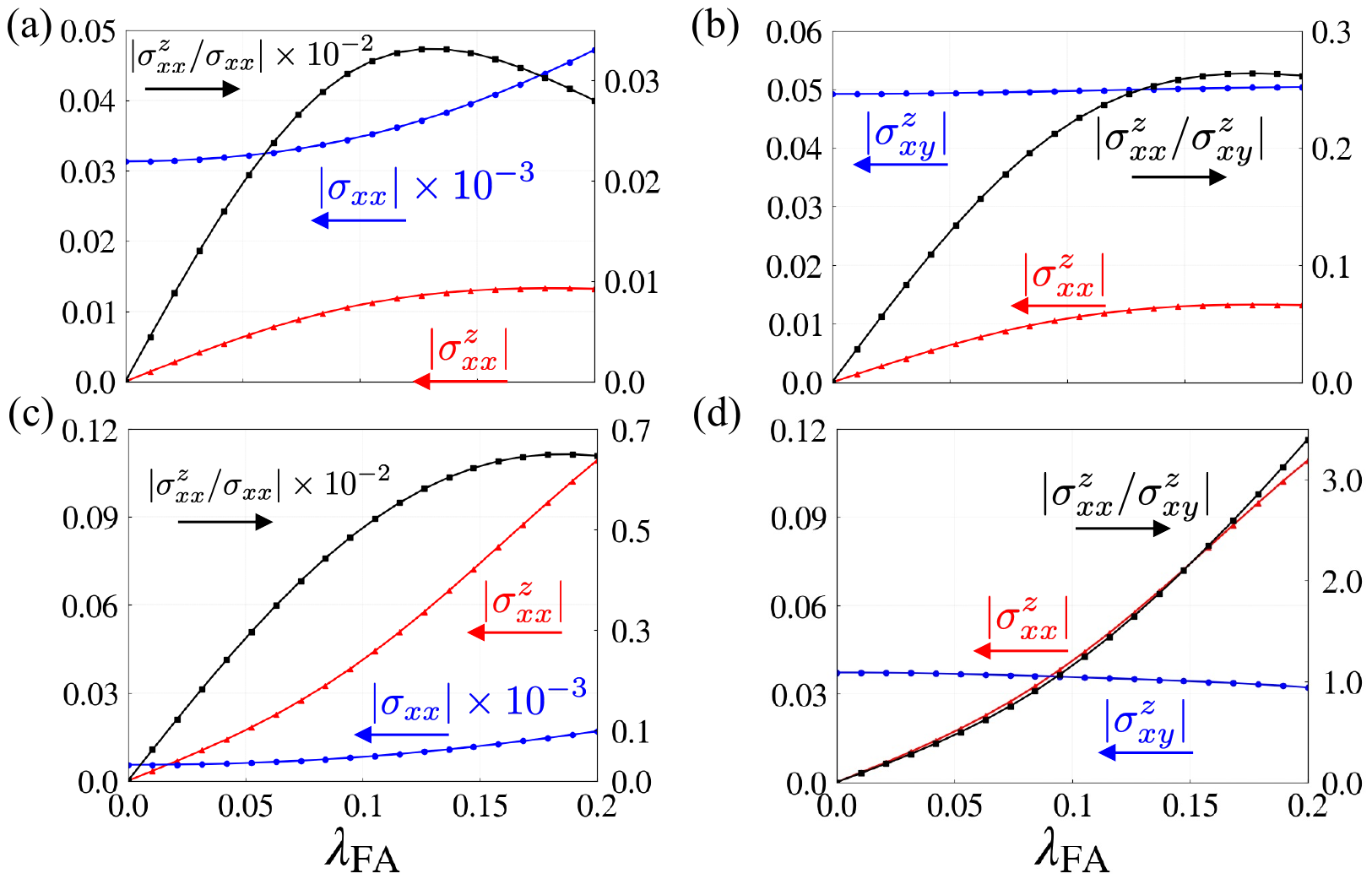}
\caption{
Comparison of the longitudinal spin conductivity, spin Hall conductivity, and longitudinal electric conductivity in the metallic regime.
(a),(b) Results for $\mu=1.5$.
(c),(d) Results for $\mu=1.05$.
}
\label{fig:cond} 
\end{figure}

In the insulating regime, the enhancement of the thermodynamic ETD has a direct transport consequence.
Equation~\eqref{eq:Gz_sigma} relates the chemical-potential derivative of $G_z$ to the intrinsic longitudinal spin conductivity, while that of $Q_0$ corresponds to the spin Hall conductivity.
Therefore, the enhanced ETD near the small band gap directly implies an enhanced longitudinal spin response.

The thermodynamic relation derived above is valid only for insulating systems.
To extend our analysis to the metallic regime, we explicitly calculate the longitudinal spin and charge conductivities.
The longitudinal electrical conductivity is evaluated using the semiclassical Boltzmann theory in the relaxation-time approximation with a phenomenological relaxation time $\tau=10^2$.

Figures~\ref{fig:cond}(a) and \ref{fig:cond}(c) show the $\lambda_{\mathrm{FA}}$ dependence of the longitudinal spin
conductivity $\sigma^{z}_{xx}$ and the longitudinal electric conductivity $\sigma_{xx}$ for $\mu=1.5$ and $1.05$, respectively.
To quantify the spin--charge conversion efficiency, we define the ratio $|\sigma^{z}_{xx}/\sigma_{xx}|$.
The conversion efficiency depends sensitively on the chemical potential and is enhanced by more than an order of magnitude near the small band gap.

Figures~\ref{fig:cond}(b) and \ref{fig:cond}(d) compare the longitudinal spin conductivity $\sigma^{z}_{xx}$ with the spin Hall conductivity $\sigma^{z}_{xy}$.
While $\sigma^{z}_{xx}$ increases monotonically with $\lambda_{\mathrm{FA}}$, reflecting its direct connection to ferroaxial order, $\sigma^{z}_{xy}$ remains nearly unchanged.
For an appropriate chemical potential, $|\sigma^{z}_{xx}|>|\sigma^{z}_{xy}|$, demonstrating that the ferroaxial longitudinal spin response can even exceed the conventional spin Hall response.

In summary, we have established a thermodynamic formulation of the ETD in periodic crystals by incorporating
the relativistic spin-induced electric polarization into the thermodynamic multipole formalism.
The resulting gauge-invariant expression provides a bulk order parameter for ferroaxial phases and directly connects its chemical-potential derivative to the intrinsic longitudinal spin conductivity in insulating systems.
Applying the formulation to a minimal ferroaxial extension of the Kane--Mele model, we have shown that the ETD emerges exclusively in the ferroaxial phase and is strongly enhanced near a small band gap, accompanied by a pronounced longitudinal spin response.
Moreover, in the metallic regime, the longitudinal spin conductivity can become comparable to, or even exceed, the conventional spin Hall conductivity.
Because the present formulation is gauge invariant and expressed entirely in terms of bulk Bloch states, it can be directly combined with first-principles calculations to quantitatively evaluate the ETD in realistic ferroaxial materials.
Our work establishes a thermodynamic framework for electric toroidicity in crystalline solids and opens a route toward quantitative identification of ferroaxial order through first-principles calculations and spin-transport measurements.

This research was supported by JSPS KAKENHI Grant Numbers JP22H00101, JP23H04869, and JP26K22277, and by JST CREST (JPMJCR23O4) and JST FOREST (JPMJFR2366).

\bibliography{main_ETD}

\end{document}

\title{Supplemental Material for ``Thermodynamic electric toroidal dipole and intrinsic longitudinal spin transport''}

\author{Taisei Yamanaka}
\author{Takumi Sato}
\author{Satoru Hayami}

\affiliation{
Graduate School of Science, Hokkaido University, Sapporo 060-0810, Japan
}

\maketitle
\subsection{Derivation of Eq.~(8)}
In this section, we follow the derivation of the thermodynamic multipoles discussed in Refs.~\cite{shi2007quantum,daido2020thermodynamic,oike2025thermodynamic,sato2026orbitalmagneticoctupolecrystalline}.
We start from the correlation function $\chi_{N, P^{\mathrm{spin}}_i}(\vb*{q}, \omega)$, giving in Eq.~(7) of the main text. 
Letting $X(\vb*{q}) = \lim_{\delta \to 0} \chi_{N, P^{\mathrm{spin}}_i}(\vb*{q}, \omega)$, 
\begin{align}
  X(\vb*{q}) = \xi \int \frac{d^dk}{(2\pi)^d} \sum_{n,m}^{} \bra{n\vb*{k}_-} \ket{m\vb*{k}_+} \bra{m\vb*{k}_+} \frac{1}{2}\left(\hat{\tilde{r}}_k (\vb*{k}_+) + \hat{\tilde{r}}_k (\vb*{k}_-) \right) \ket{n\vb*{k}_-}
  \frac{f_{n\vb*{k}_-} - f_{m\vb*{k}_+}}{\epsilon_{n\vb*{k}_-} - \epsilon_{m\vb*{k}_+}}.
\end{align}
In the case of $n\neq m$, expanding each component to first order in $q$, we obtain
\begin{align}
  \bra{n\vb*{k}_-} \ket{m\vb*{k}_+} &\simeq q_j \bra{n} \ket{\partial_j m}, \\
  \bra{m\vb*{k}_+} \frac{1}{2}\left(\hat{\tilde{r}}_k (\vb*{k}_+) + \hat{\tilde{r}}_k (\vb*{k}_-) \right)\ket{n\vb*{k}_-} &\simeq \bra{m}\hat{\tilde{r}}_i\ket{n} + \frac{q_j}{2}\left( \bra{\partial_j m}\hat{\tilde{r}}_i \ket{n} - \bra{m}\hat{\tilde{r}}_i\ket{\partial_j n} \right), \\
  \frac{f_{n\vb*{k}_-} - f_{m\vb*{k}_+}}{\epsilon_{n\vb*{k}_-} - \epsilon_{m\vb*{k}_+}}
  &\simeq \frac{f_{nm}}{\epsilon_{nm}} - \frac{q_j}{2} \frac{1}{\epsilon_{nm}} \left(\partial_j \tilde{f}_{nm} - \frac{\partial_j \tilde{\epsilon}_{nm}}{\epsilon_{nm}}f_{nm} \right),
\end{align}
where we use the shorthand notation on the right-hand side: $\ket{m\vb*{k}} = \ket{m}$, $\epsilon_{nm} = \epsilon_{n\vb*{k}} - \epsilon_{m\vb*{k}}$, $\tilde{\epsilon}_{nm} = \epsilon_{n\vb*{k}} + \epsilon_{m\vb*{k}}$, $f_{nm} = f_{n\vb*{k}}-f_{m\vb*{k}}$, $\tilde{f}_{nm} = f_{n\vb*{k}}+f_{m\vb*{k}}$.
From these equations, we obtain the contribution for $n \neq m$ as
\begin{align}\label{eq:inter}
  X^{(n \neq m)}(\vb*{q}) 
  &= q_j \xi \int \frac{d^dk}{(2\pi)^d} \sum_{n, m}^{n \neq m} \bra{n}\ket{\partial_j m} \bra{m}\hat{\tilde{r}}_i\ket{n} \frac{f_{nm}}{\epsilon_{nm}} \nonumber\\
  &= i q_j \xi \int \frac{d^dk}{(2\pi)^d} \sum_{n, m}^{n \neq m} 2\Im\left[\frac{\bra{n}\hat{\tilde{r}}_i\ket{m}\bra{m}\hat{v}_j\ket{n}}{\epsilon^2_{nm}}\right]f_{n\vb*{k}} \nonumber\\
  &= -i q_j \xi \int \frac{d^dk}{(2\pi)^d} \sum_{n} \Omega^{\tilde{r}_i, v_j}_{n\vb*{k}} f_{n\vb*{k}}
\end{align}
Here, we use $\bra{n}\ket{\partial_j m} = \bra{n}\hat{v}_j\ket{m} / \epsilon_{mn}$.
In the case of $n = m$,
\begin{align}
  \bra{n\vb*{k}_-} \ket{n\vb*{k}_+} &\simeq 1 + q_j\bra{n} \ket{\partial_j n} \nonumber,\\
  \bra{n\vb*{k}_+} \frac{1}{2}\left(\hat{\tilde{r}}_k (\vb*{k}_+) + \hat{\tilde{r}}_k (\vb*{k}_-) \right)\ket{n\vb*{k}_-} &\simeq \bra{n} \hat{\tilde{r}}_i \ket{n} + \frac{q_j}{2}\left( \bra{\partial_j n}\hat{\tilde{r}}_i \ket{n} - \bra{n}\hat{\tilde{r}}_i\ket{\partial_j n} \right), \nonumber\\
  \frac{f_{n\vb*{k}_-} - f_{n\vb*{k}_+}}{\epsilon_{n\vb*{k}_-} - \epsilon_{n\vb*{k}_+}}
  &\simeq f'_{n\vb*{k}} .
\end{align}
Here, $f'_{n\vb*{k}} = \partial f_{n\vb*{k}} / \partial\epsilon_{n\vb*{k}}$.
From these equations, we obtain the contribution for $n = m$ as
\begin{align}\label{eq:intra}
  X^{(n = m)}(\vb*{q}) 
  &= q_j \xi \int \frac{d^dk}{(2\pi)^d} \sum_{n}^{} \left\{\bra{n} \ket{\partial_j n}\bra{n} \hat{\tilde{r}}_i \ket{n} + \frac{1}{2}\left( \bra{\partial_j n}\hat{\tilde{r}}_i \ket{n} - \bra{n}\hat{\tilde{r}}_i\ket{\partial_j n} \right)\right\} f'_{n\vb*{k}} \nonumber\\
  &= q_j \xi \int \frac{d^dk}{(2\pi)^d} \sum_{n, m}^{n \neq m} \frac{1}{2}\left( \bra{\partial_j n}\ket{m} \bra{m}\hat{\tilde{r}}_i \ket{n} - \bra{n}\hat{\tilde{r}}_i\ket{m} \bra{m}\ket{\partial_j n} \right) f'_{n\vb*{k}} \nonumber\\
  &= -i q_j \xi \int \frac{d^dk}{(2\pi)^d} \sum_{n, m}^{n \neq m} \Im\left[\frac{\bra{n}\hat{\tilde{r}}_i\ket{m}\bra{m}\hat{v}_j\ket{n}}{\epsilon_{nm}}\right]f'_{n\vb*{k}} \nonumber\\
  &= -i q_j \xi \int \frac{d^dk}{(2\pi)^d} \sum_{n} m^{\tilde{r}_i, v_j}_{n\vb*{k}}f'_{n\vb*{k}}
\end{align}
Combining \Eq{eq:intra} and \Eq{eq:inter}, we obtain
\begin{align}
  \frac{\partial Q_{ij}}{\partial \mu} 
  &= -i\lim_{\vb*{q} \to 0} \partial_{q_j} X(\vb*{q}) \nonumber\\
  &= -\xi\int_{}^{}\frac{d^dk}{(2\pi)^d} \sum_{n}^{} 
  \Bigg[ \Omega^{\tilde{r}_i, v_j}_{n\vb*{k}} f_{n\vb*{k}} + m^{\tilde{r}_i, v_j}_{n\vb*{k}}f'_{n\vb*{k}} \Bigg].
\end{align}
Integrating over the chemical potential yields
\begin{align}\label{eq:EQ}
  Q_{ij} 
  &= \int_{-\infty}^{\mu} \frac{\partial Q_{ij}}{\partial \mu'} d\mu' \nonumber\\
  &= \xi\int_{}^{}\frac{d^dk}{(2\pi)^d} \sum_{n}^{} 
  \Bigg[ \Omega^{\tilde{r}_i, v_j}_{n\vb*{k}} \mathcal{G}_{n\vb*{k}} + m^{\tilde{r}_i, v_j}_{n\vb*{k}}f_{n\vb*{k}} \Bigg].
\end{align}
This corresponds to Eq.~(8) 
in the main text.

\subsection{Contribution from the M-point band gap}
\begin{figure}[H]
\includegraphics[width=\linewidth]{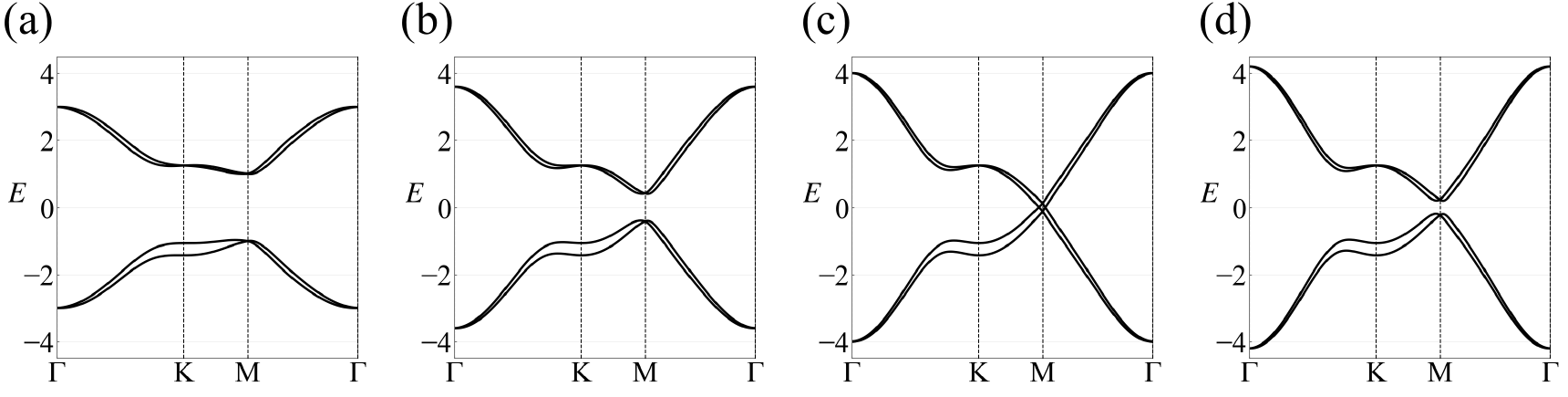}
\caption{
The band structure of the model at $t=1$, $\lambda_{\text{SO}}=0.24$, $\lambda_{\text{R}}=0.06$ and $\lambda_{\text{FA}}=0.1$: (a) $t'=0$, (b) $t'=0.2$, (c) $t'=1/3$, (d) $t'=0.4$.
}
\label{fig:band_t_dep} 
\end{figure}

To clarify the role of the M-point band gap in the enhancement of the thermodynamic ETD and the intrinsic longitudinal spin conductivity, we introduce an additional third-nearest-neighbor hopping term,
\begin{align}
  H_{\mathrm{t,3rd}} = -t'\sum_{i}^{} \sum_{\sigma}^{}e^{-i\vb*{k} \cdot \vb*{\delta}'_i} c^\dag_{\text{A} \sigma \vb*{k}}c_{\text{B} \sigma \vb*{k}} + \text{h.c.},
\end{align}
where, $\vb*{\delta}'_1 = \vb*{\delta}_1+\vb*{\delta}_2-\vb*{\delta}_3, \vb*{\delta}'_2 = \vb*{\delta}_1-\vb*{\delta}_2+\vb*{\delta}_3, \vb*{\delta}'_3 = -\vb*{\delta}_1+\vb*{\delta}_2 +\vb*{\delta}_3$.
The M-point band gap is controlled by this term.
Figure~\ref{fig:band_t_dep} shows the band structures for several $t'$: (a) $t'=0$, (b) $t'=0.2$, (c) $t'=1/3$, (d) $t'=0.4$. 
The parameters are set to $t=1$, $\lambda_{\text{SO}}=0.24$, $\lambda_{\text{R}}=0.06$ and $\lambda_{\text{FA}}=0.1$.
We find that as $t'$ increases, the overall band shape remains largely unchanged, and the gap at the M-point closes at $t' = 1/3$
and then reopens.

\begin{figure}[H]
\includegraphics[width=\linewidth]{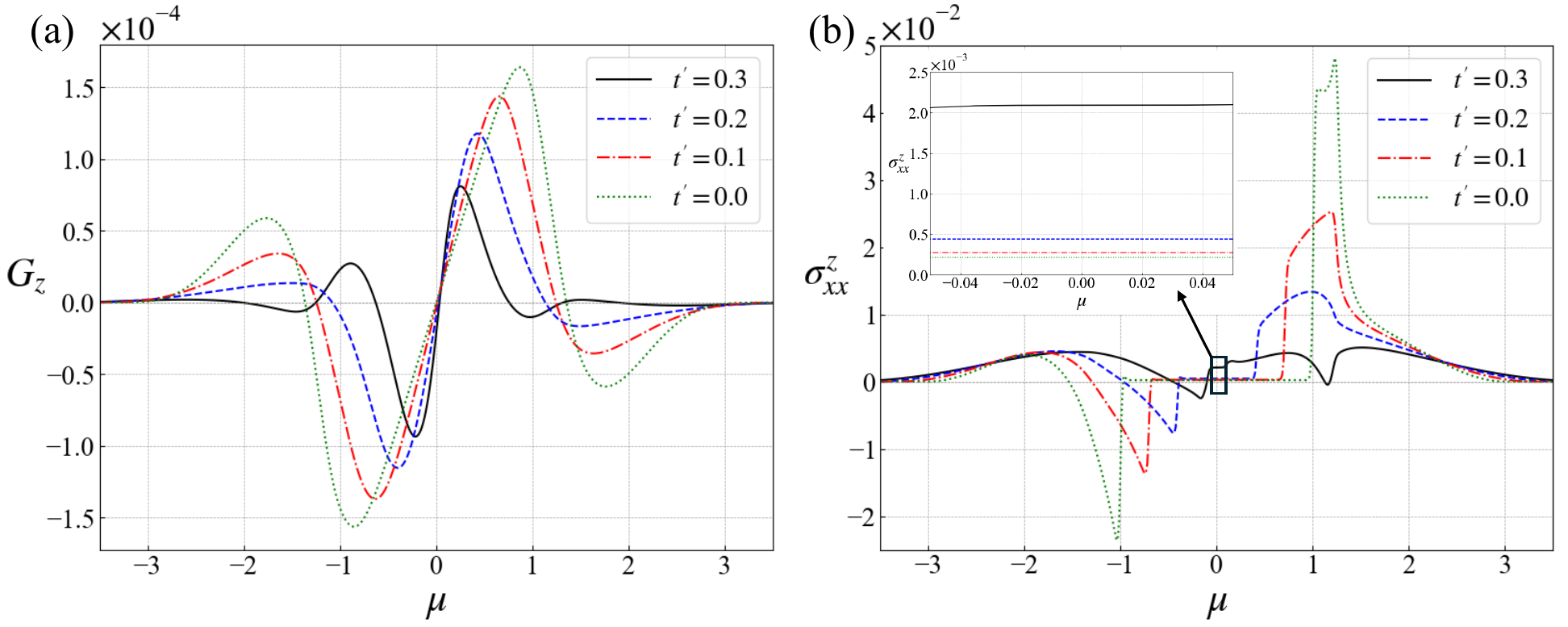}
\caption{
(a) Chemical potential dependence of thermodynamic ETD $G_z$ and (b) the longitudinal spin conductivity $\sigma^z_{xx}$ for $t' = 0.0$, $0.1$, $0.2$, and $0.3$, respectively.
}
\label{fig:t_dep} 
\end{figure}

Figures~\ref{fig:t_dep}(a) and \ref{fig:t_dep}(b) show the chemical potential dependence of thermodynamic $G_z$ and the longitudinal spin conductivity $\sigma^z_{xx}$ for $t' = 0.0$, $0.1$, $0.2$, and $0.3$, respectively.
Both the slope of $G_z$ and the magnitude of $\sigma^{z}_{xx}$ in the insulating regime increase as the M-point band gap decreases, i.e., as $t'$ approaches $1/3$.
Since the system remains a topological insulator before the gap closes, the slope of $Q_0$ and the spin Hall conductivity $\sigma^{z}_{xy}$ remain unchanged. 
In contrast, in the metallic regime, the magnitudes of $G_z$ and $\sigma^{z}_{xx}$ tend to decrease with increasing $t'$. 
This trend is consistent with the increased bandwidth away from the M point, which enlarges the energy denominators in Eq.~(10)
of the main text.

\subsection{Essential parameters for the longitudinal spin conductivity}
Following the systematic analysis developed in Ref.~\cite{oiwa2022systematic}, we identify the model parameters required for a finite intrinsic longitudinal spin conductivity.
To this end, we evaluate
\begin{align}\label{eq:essential}
  \Im\left[ \Gamma^{ij}_{xx}(z)\right] = \sum_{\vb*{k}}^{} \Tr\left[ \hat{j}^{z}_x \mathcal{H}^{i} \hat{v}_x \mathcal{H}^{j}\right].
\end{align}
Keeping the lowest-order and second-lowest-order contributions to \Eq{eq:essential} yields 
\begin{align}
  \Im\left[ \Gamma^{12}_{xx}(z)\right] &= 3 \lambda_{\text{FA}} \lambda_{\text{R}}^2  \left(-\lambda_{\text{R}}^2 + 2t^2 \right), \nonumber \\
  \Im\left[ \Gamma^{23}_{xx}(z)\right] &= 3 \lambda_{\text{FA}} \lambda_{\text{R}}^2 \left(6 \lambda_{\text{R}}^4 + 4 \lambda_{\text{R}}^2 \lambda_{\text{SO}}^{2} - 18 \lambda_{\text{R}}^{2} t^{2} 
+ 10 \lambda_{\text{R}}^{2} \lambda_{\text{FA}}^{2} + 12 \lambda_{\text{SO}}^{2} t^{2} + 3 t^{4} + 150 t^{2} \lambda_{\text{FA}}^{2}\right),
\end{align}
From these results, $\sigma^z_{xx}$ can be expreesed as
\begin{align}
  \sigma^z_{xx} = 3 \lambda_{\text{FA}} \lambda_{\text{R}}^2 F(t, \lambda_{\text{SO}}, \lambda_{\text{R}}, \lambda_{\text{FA}}),
\end{align}
where, $F$ is a function of the parameters $(t, \lambda_{\text{SO}}, \lambda_{\text{R}}, \lambda_{\text{FA}})$.
Therefore, we find that a finite intrinsic longitudinal spin conductivity requires both $\lambda_{\text{FA}}$ and $\lambda_{\text{R}}$ to be finite.
\bibliographystyle{apsrev4-2}
\bibliography{main_ETD}